\documentclass[a4paper,twocolumn,superscriptaddress,prl, aps, floatfix ]{revtex4-1}
\usepackage[pdftex]{graphicx}
\usepackage{latexsym,amsmath}
\usepackage[english]{babel}
\usepackage[pdftex]{color}
\usepackage{hyperref}
\usepackage{microtype}
\usepackage{comment}

\usepackage{amssymb}
\usepackage{amsmath}
\usepackage{amsfonts}
\usepackage{color}
\usepackage{multirow,tabularx}

\usepackage{scalerel}

\newcommand\reallywidehat[1]{\arraycolsep=0pt\relax%
\begin{array}{c}
\stretchto{
  \scaleto{
    \scalerel*[\widthof{\ensuremath{#1}}]{\kern-.5pt\bigwedge\kern-.5pt}
    {\rule[-\textheight/2]{1ex}{\textheight}} %WIDTH-LIMITED BIG WEDGE
  }{\textheight} % 
}{0.5ex}\\           % THIS SQUEEZES THE WEDGE TO 0.5ex HEIGHT
#1\\                 % THIS STACKS THE WEDGE ATOP THE ARGUMENT
\rule{-1ex}{0ex}
\end{array}
}

\newcommand{\av}[1]{\langle #1 \rangle}

\newcommand{\FigPath}{.}

\begin{document}
\title{Equivalence between non-Markovian and Markovian dynamics\\ in epidemic spreading processes}

\author{Michele \surname{Starnini}}
\thanks{Corresponding author: michele.starnini@gmail.com}
\affiliation{Departament de F\'isica de la Mat\`eria Condensada, Universitat de Barcelona, Mart\'i i Franqu\`es 1, E-08028 Barcelona, Spain}
\affiliation{Universitat de Barcelona Institute of Complex Systems (UBICS), Universitat de Barcelona, Barcelona, Spain}

\author{James P.~\surname{Gleeson}}
\affiliation{MACSI, Department of Mathematics and Statistics, University of Limerick, Ireland}

\author{Mari\'an \surname{Bogu\~n\'a}}
\thanks{Corresponding author: marian.boguna@ub.edu}
\affiliation{Departament de F\'isica de la Mat\`eria Condensada, Universitat de Barcelona, Mart\'i i Franqu\`es 1, E-08028 Barcelona, Spain}
\affiliation{Universitat de Barcelona Institute of Complex Systems (UBICS), Universitat de Barcelona, Barcelona, Spain}

\begin{abstract}
A general formalism is introduced to allow the steady state of non-Markovian processes on networks to be reduced to equivalent Markovian processes on the same substrates. The example of an epidemic spreading process is considered in detail, where all the non-Markovian aspects are shown to be captured within a single parameter, the effective infection rate. Remarkably, this result is independent of the topology of the underlying network, as demonstrated by numerical simulations on two-dimensional lattices and various types of random networks. Furthermore, an analytic approximation for the effective infection rate is introduced, which enables the calculation of the critical point and of the critical exponents for the non-Markovian dynamics.
\end{abstract}

\maketitle

Modeling the stochastic dynamics that occur in many natural and technological systems has long depended on the Markovian assumption. In a Markov process, the probabilities of the occurrence of future events depend only on the present state of the system, being independent of the prior history. This memoryless property implies that such dynamics
can be described by Poisson processes with fixed rates, which are characterized by an exponential distribution of the inter-event time between consecutive events~\cite{opac-b1080435}. The mathematical tractability of Markov processes enables great simplifications in problem formulation, leading to spectacular successes in the description of many dynamical processes unfolding on networks~\cite{barratbook} and in other complex systems.

The dominance of the Markovian modeling framework has recently been challenged by the  increasing availability of time-resolved data on different kind of interactions, ranging from human activity patterns, including communication and mobility~\cite{Barabasi:2005uq,Oliveira:2005fk,Gonzalez:2008fk}, to natural phenomena~\cite{PhysRevLett.92.108501, Wheatland1998}, biological processes~\cite{journals/biosystems/KemuriyamaOSMTKN10}, and biochemical reactions~\cite{Bratsun11102005}. These empirical observations have revealed correlated sequences of events with heavy tailed interevent time distributions~\cite{Karsai:2012aa}, a clear signature that the homogeneous temporal process description is inadequate and that non-Markovian dynamics lie at the core of such interactions.

Meanwhile, the interest in non-Markovian dynamical processes within the complex systems community has blossomed,
from the points of view of both mathematical modeling~\cite{Moinet2015, Karsai:2014aa, Garcia-Perez:2015aa,Jo2014,Kiss:2015aa} and numerical simulation~\cite{van_mieghem_non-markovian_2013, boguna_simulating_2013}.
Particular attention has been devoted to epidemic spreading on complex networks,
representing the diffusion of information or disease in a population~\cite{RevModPhys.87.925}.
Recently, it has been shown that a non-Markovian infection dynamics dramatically alter the Susceptible-Infected-Susceptible (SIS) spreading process~\cite{van_mieghem_non-markovian_2013, PVM_nonMarkovianSIS_NIMFA_2013}.
non-Markovian effects are now known to give qualitatively new behavior in information spreading, e.g., on social networks,
as revealed by measurements of inter-event times for email responses~\cite{Iribarren09,Iribarren11}
and retweets on Twitter~\cite{Lerman16}.
In the context of epidemiology, the non-Markovian assumption is particularly relevant,
as empirical measurements of real diseases---smallpox, measles, ebola---indicate that
the distribution of infectious periods is far from being exponential~\cite{BAILEY01061956,Eichner15072003,nishiura2006,Chowell:2014aa}.

In this Letter, we consider the non-Markovian SIS epidemic model and
show that its steady-state is equivalent to a Markovian one with an effective infection rate $\lambda_{eff}$, thus 
encoding all the non-Markovian effects into a single parameter.
Interestingly, this result is independent of the underlying network topology.
Our mathematical formalism demonstrates the existence of the effective rate $\lambda_{eff}$,
allowing us to compute it by means of numerical simulations,
and enables us to derive an approximate analytic expression $\lambda_{app}$,
 in very good agreement with $\lambda_{eff}$.
The approximate value $\lambda_{app}$ is expected to converge to $\lambda_{eff}$ close to the epidemic threshold, and therefore the critical point 
and the set of critical exponents of the non-Markovian SIS dynamics, when expressed in terms of $\lambda_{app}$, are the same of those of the Markovian case,
as we show by means of a finite size scaling analysis.
It is worth  remarking that our formalism is not restricted to the SIS model and
can be easily extended to any non-Markovian dynamics with a finite set of discrete states,
allowing the determination of the extent to which such dynamics can be reduced to a Markovian equivalent 
 (with redefined parameters) or whether the non-Markovian dynamics are fundamentally different.

Let us consider an undirected and unweighted network topology defined by an adjacency matrix $a_{ij}$, with $i,j=1,\cdots,N$,
and a general, non-Markovian SIS dynamics running on top.
In this model, nodes exist in either of two states, susceptible or infected.
Infected nodes decay spontaneously to the susceptible state after a random time $t$ distributed as $\psi_R(t)$,
 that is, recovery from the illness does not confer any long lasting immunity,
 a characteristic present in some sexually transmitted diseases~\cite{Keeling07book}.
 Susceptible nodes may become infected upon contact with infected neighbors,
and we assume that each infectious (or active) link, connecting an infected node with a susceptible one,
 hosts statistically independent stochastic infection processes,
 each one controlled by the same interevent distribution $\psi_I(t)$.
In an active link isolated from the rest of the system,
 the susceptible node becomes infected after a random time $t$ has 
 elapsed since the infection was initiated, with $t$ distributed as $\psi_I(t)$.
If a susceptible node is connected to more than one infected neighbor, infection processes take place independently along each infectious link.

Distributions $\psi_R(t)$ and $\psi_I(t)$ allow us to evaluate the (time-dependent) hazard rates,
defined as the probability per unit of time that, given that the event did not take place by a time $t$ since the process was initiated, it takes place in the time interval between $t$ and $t+dt$~\cite{renewal}. The recovery and infection hazard rates are defined as $\delta(t)= \psi_R(t) / \Psi_R(t)$ and $\lambda(t) = \psi_I(t) / \Psi_I(t)$, where $\Psi_R(t)$ and $\Psi_I(t)$ are the corresponding survival probabilities, that is, the probability that a given event takes a time longer than $t$. When temporal processes follow Poisson (Markovian) statistics, both distributions are exponential and the corresponding hazard rates are constants.

The SIS dynamics can be fully described by a set of binary stochastic processes $\{ n_i(t)\}$, $i=1,\cdots,N$; defined as
$n_i(t)=1$ if node $i$ is infected at time $t$ and zero if it is susceptible. The exact stochastic evolution of these processes can be written as
\begin{equation}
n_i(t+dt)=n_i(t) \xi_i(t,dt)+[1-n_i(t)] \eta_i(t,dt).
\label{eq:master}
\end{equation}
In this equation, the first term in the sum of the right hand side is different from zero only when node $i$ is infected and accounts for its recovery during the time interval $(t,t+dt)$. To achieve this, the stochastic process $\xi_{i}(t,dt)$ is defined to be equal to zero with probability
$dt\delta[t_i(t)]$ and one otherwise, where $t_i(t)$ is the time elapsed, at time $t$, since node $i$ became infected. Similarly, the second term in the sum of the right hand side of Eq.~\eqref{eq:master} accounts for the infection of susceptible node $i$ by one of its infected neighbors during the time interval $(t,t+dt)$. The the stochastic process $\eta_{i}(t,dt)$ is defined to be equal to one with probability $dt \sum_j a_{ij} n_j(t) \lambda[\tau_{ji}(t)]$
and zero otherwise, where $\tau_{ji}(t)$ is the time elapsed since the infection process of node $j$ to node $i$ started. Note that we implicitly assume that each infected neighbor defines a statistically independent random process so that the total infection hazard rate is simply the sum of the infection hazard rates of each individual process. Note also that in this formulation $t_i(t)$ and $\tau_{ji}(t)$ are themselves stochastic processes.

The average of Eq.~\eqref{eq:master}, first conditioned to the knowledge of the stochastic processes $\{n_i,t_i,\tau_{ji}\}$ at time $t$, and then over the unconditional values, allows us to write the following differential equation for the probability of node $i$ to be infected at time $t$, $\rho_i(t)\equiv \langle n_i(t) \rangle$~\footnote{Note that with this definition, the prevalence of the disease at time $t$ is simply given by $\rho(t)=N^{-1} \sum_{i=1}^N \rho_i(t)$.}
\begin{equation}
\dot{\rho}_i(t) = -\langle n_i(t)\delta[t_i(t)] \rangle+\sum_{j=1}^N a_{ij} \langle [1-n_i(t)]n_j(t) \lambda[\tau_{ji}(t)] \rangle.
\label{eq:master2}
\end{equation}
The first term in Eq.~\eqref{eq:master2} can be rewritten as (see Supplementary Information, SI)
\begin{equation}
\langle n_i(t)\delta(t_i(t)) \rangle = \rho_i(t) \langle \delta[t_i(t)] |n_i=1 \rangle.
\end{equation}
In the limit $t \rightarrow \infty$, the only information we have about $t_i$, given that node $i$ is infected, is that
the recovery time of node $i$ after infection is longer than $t_i$. This implies that the probability density of $t_i$ is given by $\Psi_R(t_i)/\langle t_R \rangle$, where $\langle t_R \rangle$ is the average recovery time~\cite{renewal}. By combining this result with the form of the recovery hazard rate, we can write
\begin{equation}
\label{eq:delta_ti}
\lim_{t\rightarrow \infty} \langle n_i(t)\delta(t_i(t)) \rangle = \rho_i^{st} \int_0^\infty  \frac{\Psi_R(t_i)}{\langle t_R \rangle} \delta(t_i) dt_i = \frac{\rho_i^{st}}{\langle t_R \rangle} ,
\end{equation}
where we have defined $\rho_i^{st}=\lim_{t\rightarrow \infty} \rho_i(t)$. Similarly, the terms on the right hand side of Eq.~\eqref{eq:master2} can be written as
\begin{equation}
\begin{array}{l}
\langle [1-n_i(t)]n_j(t) \lambda[\tau_{ji}(t)] \rangle=\\[0.3cm]
= \langle [1-n_i(t)]n_j(t) \rangle \langle \lambda[\tau_{ji}(t)] |n_i=0,n_j=1 \rangle.
 \end{array}
\end{equation}
From this equation, we observe that the evolution of the density $\rho_i(t)$ depends on the evolution of two-point correlation functions, $\rho_{ij}(t)=\langle n_i(t) n_j(t) \rangle$, that appear in the second term of Eq.~\eqref{eq:master2}. Using similar arguments to those used to derive Eq.~\eqref{eq:master2}, we can write an exact differential equation for
the $n$-point correlation function $\rho_{i_1\cdots i_n}$ (see SI):
\begin{equation}
\dot{\rho}_{i_1\cdots i_n}  = \sum_{i \in \mathcal{I}} \langle  \left[ - \delta(t_i) n_i +
 (1-n_i) \sum_{j=1}^N a_{ij} n_j \lambda(\tau_{ji}) \right] \prod_{k \in \mathcal{I}_i} n_k \rangle,
\label{eq:master3}
\end{equation}
where we omit the dependence on $t$ for brevity and we define the sets of nodes $\mathcal{I} \equiv \{i_1, i_2, \cdots i_n  \}$ and $\mathcal{I}_i \equiv \mathcal{I} \setminus i$.
Eq.~\eqref{eq:master3} can be written as
\begin{equation}
\begin{array}{rcl}
\dot{ \rho}_{i_1\cdots i_n}  &=&  - \rho_{i_1\cdots i_n}  \sum_{i \in \mathcal{I}} \tilde{\delta}_i + \\[0.4cm]
&+ & \sum_{i \in \mathcal{I}}\sum_{j=1}^N a_{ij}\tilde{\lambda}_{ji}[\rho_{i_1\cdots j \cdots i_n}-\rho_{i_1\cdots i_{n} j}]
\end{array}
\label{eq:npoint_corr}
\end{equation}
where $\rho_{i_1\cdots j \cdots i_n}$ and $\rho_{i_1\cdots i_{n} j}$ are the $n$ and $(n+1)$-point correlation functions of the sets $\mathcal{I}_i \cup \{j\}$ and $\mathcal{I} \cup \{j\}$, respectively, and where we have also defined
\begin{equation}
 \tilde{\delta}_i(t)  \equiv  \langle \delta[t_i (t)] | \{n_j = 1, j \in \mathcal{I} \}\rangle
 \label{eq:delta_cond}
\end{equation}
and
 \begin{equation}
  \tilde{\lambda}_{ji}(t)  \equiv   \langle \lambda[\tau_{ji} (t)] | n_i = 0, n_j=1, \{n_k = 1, k \in \mathcal{I}_i \} \rangle.
  \label{eq:lambda_cond}
\end{equation}
Equations~\eqref{eq:npoint_corr},~\eqref{eq:delta_cond}, and~\eqref{eq:lambda_cond} are the central result of our paper as they fully describe the dynamics of the epidemic. However, what makes our formulation interesting is the fact that all the non-Markovian effects of the dynamics are encoded in the terms $\tilde{\delta}_i$ and $\tilde{\lambda}_{ji}$. As we shall show later, under certain conditions these parameters take constant values independent of the nodes, that is, $\tilde{\delta}_i=\tilde{\delta}$ and $\tilde{\lambda}_{ji}=\tilde{\lambda}$. In this case, the dynamics, even if strongly non-Markovian, can be described by a Markovian one on the same network, using effective parameters $\tilde{\delta}$ and $\tilde{\lambda}$. In this way, the considerable complexity of the non-Markovian effects is reduced to the evaluation of such effective parameters.
\begin{figure}[tbp]
  \begin{center}
    \includegraphics[width=6cm,angle=0]{\FigPath/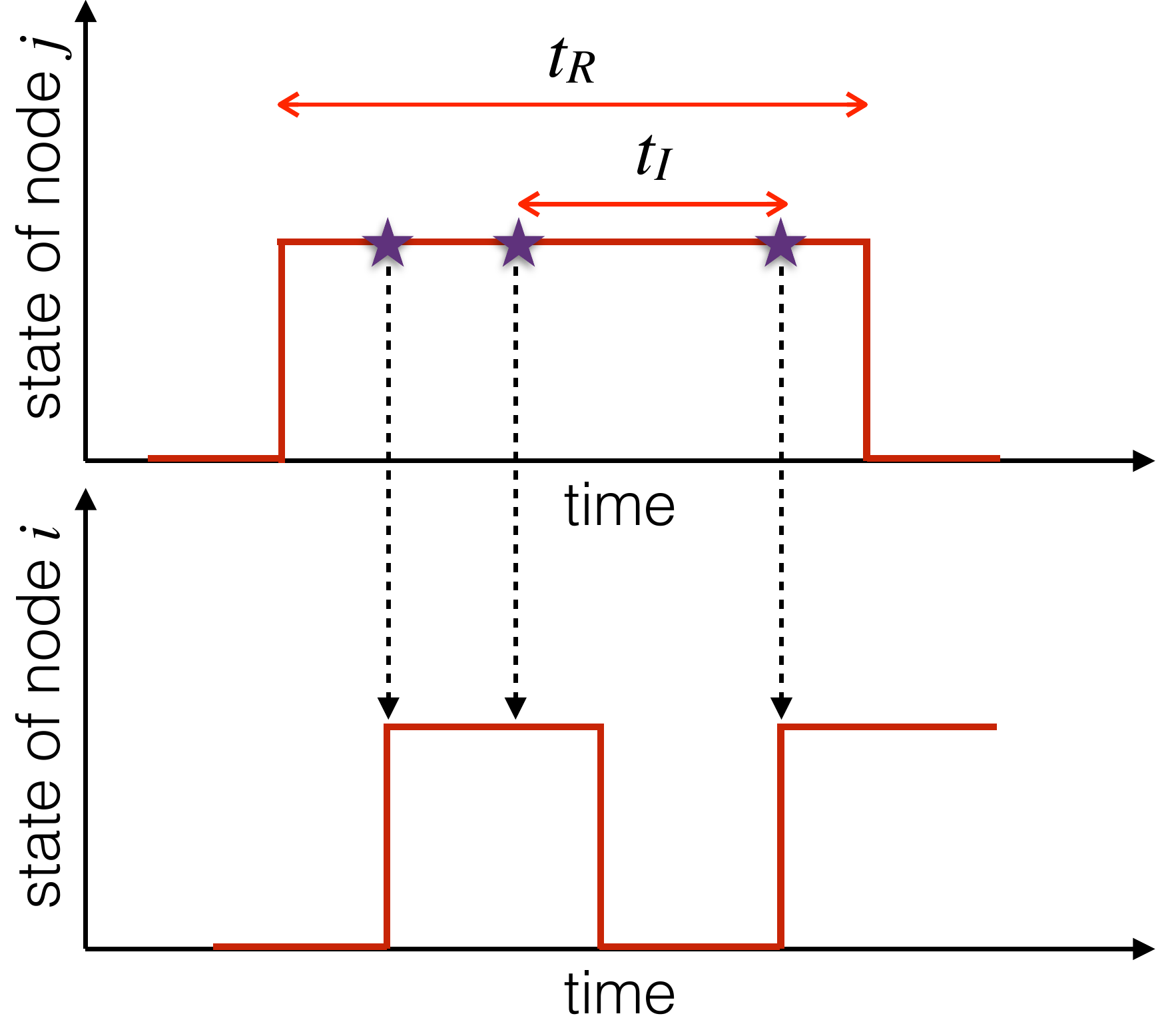}
  \end{center}
  \vspace{-0.6cm}
  \caption{Sketch of the infection mechanism from node $j$ to node $i$.
  Infection events triggered by node $j$ (represented by stars in the figure) are ineffective
  if node $i$ is already infected.
}
  \label{sketch}
\end{figure}

To proceed further, we need to define the details of the pairwise interaction that rules the infection process.
We assume that the infection process between an infected node $j$ and a susceptible node $i$ depends on the state of node $j$ alone,
i.e. when a node $j$ becomes infected, it starts an infection process independently to each of his neighbors,
regardless of their state, according to a renewal process with inter-event time distribution $\psi_I(t)$.
One can think of this process as a series of firing events, separated by random times $t_I$,
starting when node $j$ becomes infected, so that when one such event takes place at a time that neighbor node $i$ is susceptible, node $i$ becomes  infected (see Fig.~\ref{sketch}).
We also assume that the recovery process of an infected node depends on its state alone,
i.e., when a node becomes infected, it starts a recovery process with random time $t_R$, distributed as $\psi_R(t)$.

Within this framework, the average of $\lambda[\tau_{ji}(t)]$ conditioned to the state of the system can be derived by noting that,
at time $t$, the time elapsed since the infection process of node $j$ to node $i$ started, $\tau_{ji}(t)$,
is not truly independent of the state of $i$.
That is, if the time elapsed since node $i$ has recovered is $\tau^R_i$, then it holds that $\tau_{ji} > \tau^R_i$.
Therefore, $\tau_{ji}$ depends on the state of $i$ but {\it not on the state of any other neighbor} and, consequently,
$\text{Prob}(\tau_{ji};t |  n_i = 0, n_j = 1, \{n_k=1, k \in \mathcal{I}_i\}) = \text{Prob}(\tau_{ji} ;t | n_i = 0,  n_j = 1)$
(see SI for a detailed proof).
This implies that we can then define an effective infection rate $\lambda_{eff}$ as
\begin{eqnarray}
\lambda_{eff} & \equiv & \lim_{t\rightarrow \infty} \langle \lambda[\tau_{ji}(t)] | n_i=0, n_j=1\rangle \nonumber \\
  & = & \int_0^{\infty} \phi (\tau_{ji})\lambda(\tau_{ji}) d\tau_{ji},
\label{eq:lambda_eff}
\end{eqnarray}
where $\phi (\tau_{ji}) \equiv \lim_{t\rightarrow \infty}  \text{Prob}(\tau_{ji};t | n_j=1, n_i=0 )$ is
the probability density of $\tau_{ji}$, where $\tau_{ji}$ is the time elapsed since the start of the infection process from node $j$ to node $i$, given that node $i$ is susceptible and node $j$ is infected and $\lambda[\tau_{ji}(t)]$ is averaged over all active links $i-j$ in the network.

Concerning the average of $\delta[t_i (t)]$, we note that in general, in the long time limit,
Eq.~\eqref{eq:delta_cond} does not reduce to Eq.~\eqref{eq:delta_ti},
since $\text{Prob}(t_i;t | \{n_k = 1, k \in \mathcal{I} \}) \neq \text{Prob}(t_i;t | n_i=1)$.
This is due to the fact that, especially for low-degree nodes, the time $t_i(t)$ may depend on the status of node $i$'s neighbors:
if at time $t$ node $j$ is infected and connected only to node $i$ ($k_j$=1), node $j$ must have been infected by node $i$,
therefore $t_i$ has to be larger than the time elapsed since the infection process of node $j$ started.
Thus, if the recovery process follows a non-Markovian dynamics it is not possible to define an effective parameter $\delta_{eff}$.
Therefore, hereafter we consider only the case of Markovian recovery, which implies $\tilde{\delta}_i=\tilde{\delta}=\langle t_R \rangle^{-1}$,
so that the non-Markovian SIS dynamics can be reduced to a Markovian one with parameters $\tilde{\delta}$ and $\lambda_{eff}$.

Although the probability density $ \phi (\tau_{ji})$ can be easily measured in numerical simulation,
it is too cumbersome to be computed analytically,
even in the simplest case of Markovian recovery (see SI).
 Therefore, we also evaluate an approximate effective infection rate $\lambda_{app}$.
 To do so, we first notice that $\lambda_{eff}$ can also be written as
\begin{equation}
\lambda_{eff} = \mathcal{N}^{-1} \int_0^{\infty} \psi (\tau_{ji})\lambda(\tau_{ji}) d\tau_{ji},
\label{eq:lambda_approx}
\end{equation}
where $\psi (\tau_{ji}) \equiv \lim_{t\rightarrow \infty}   \text{Prob}(\tau_{ji}, n_i=0; t | n_j=1)$ is
the joint probability that node $i$ is susceptible and the time elapsed since the last infection attempt from $j$ to $i$ is equal to $\tau_{ji}$, given that node $j$ is infected at a given observation time $t \rightarrow \infty$,
and where $\mathcal{N}$ is the normalization factor $\mathcal{N} = \int_0^\infty  \psi(\tau_{ji}) d\tau_{ji}$.
In the SI, we derive an approximate analytic expression for $\psi (\tau_{ji})$ that,
 combined with Eq.~\eqref{eq:lambda_approx}, allows us to derive the following expression for the approximate effective rate
\begin{equation}
\lambda_{app} =    \frac{  \widehat{\psi}_I ( 2 \tilde{\delta})  +
\av{k}  \widehat{\psi}_I (  \tilde{\delta})    \left[   1-  \widehat{\psi}_I ( 2 \tilde{\delta}) \right]
 \left[ \widehat{\psi}_I ( \tilde{\delta}) -1  \right]^{-1}  }
{    \left[  \av{k}-1 \right]     \widehat{\Psi}_I(\tilde{\delta}) }
\label{eq:lambda_approx_res}
\end{equation}
where $\widehat{\psi}_I (u) \equiv \mathcal{L} \{ \psi_I(t) \}$ and $\widehat{\Psi}_I (u) \equiv \mathcal{L} \{ \Psi_I(t) \}$ are the Laplace transforms of $\psi_I(t)$ and $\Psi_I(t)$, respectively, and $\av{k}$ is the average degree of the network substrate.

We check the validity of the effective infection rates,
$\lambda_{eff}$ and $\lambda_{app}$,
by means of extensive numerical simulations of the non-Markovian SIS dynamics,
see SI.
We consider a Poissonian (Markovian) recovery process with rate $\tilde{\delta}$
and an infection process with a Weibull inter-event time distribution, that is,
\begin{equation}
\label{eq:psi_I}
\psi_I(t) = \frac{\alpha_I}{b}\left( \frac{t}{b} \right)^{\alpha_I-1} e^{-\left( t/b \right)^{\alpha_I}}, \quad \psi_R(t) =  \tilde{\delta}e^{- \tilde{\delta}t}
\end{equation}
with parameter $\alpha_I$ controlling the power-law start and tail of the infection inter-event time distribution.
We choose $b = \av{t_I} \left[ \Gamma(1+1/\alpha_I) \right)  ]^{-1}$,
so that $\av{t_I}$ is the average infection time.
Hereafter, and without loss of generality, we set the time scale to $\tilde{\delta}=1$.
 Once the system has reached its steady state, we evaluate $\lambda_{eff}$
 by selecting random time instants along the process.
 For each time instant, we select all active links, measure the corresponding values of $\tau_{ji}$,
  and calculate $\lambda_{eff}$ as the average of the hazard rates $\lambda(\tau_{ji})=\frac{\alpha_I}{b}\left( \frac{\tau_{ji}}{b} \right)^{\alpha_I-1}$.
  The approximate infection rate $\lambda_{app}$ is calculated from Eq.~(\ref{eq:lambda_approx_res})  by integrating numerically
  the Laplace transforms  $ \widehat{\psi}_I(\tilde{\delta})$ and $ \widehat{\Psi}_I(\tilde{\delta})$ with $\tilde{\delta}=1$.

\begin{figure}[tbp]
  \begin{center}
    \includegraphics[width=\linewidth,angle=0]{\FigPath/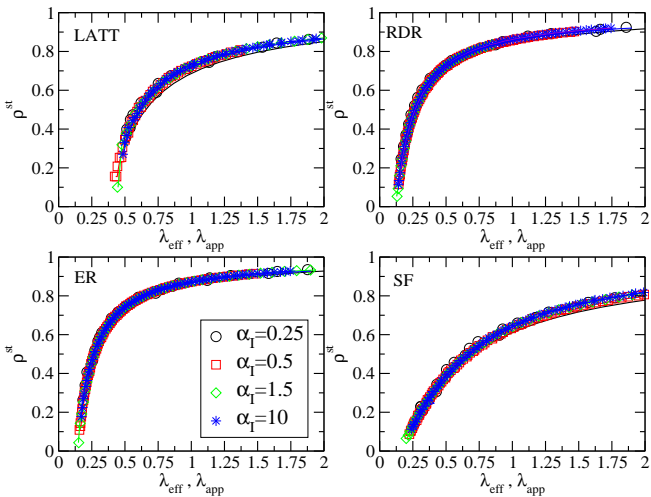}
  \end{center}
  \caption{
Steady-state prevalence $\rho^{st}$ as a function of the effective infection rate,
 	for different values of the exponent $\alpha_I$ controlling the interevent time infection distribution and different network substrates.
 	Symbols represent the effective infection rate  $\lambda_{eff}$, extracted by numerical simulations,
 	continuous lines represent the approximate rate  $\lambda_{app}$.}
  \label{fig:panel_lambda_exact}
\end{figure}

\begin{table}[tb]
\begin{ruledtabular}
    \begin{tabular}{c||ccccc}
  Topology  &  $\alpha_I$ & $\lambda_c$ &  $\beta$ & $\nu_{\perp}$ &  $\delta$  \\ \hline 
 \multirow{3}{*}{Lattice}  & 0.5 & 0.4194 & 0.596 & 0.739 & 0.453 \\ 
                                         & 2.0 & 0.4200 & 0.594 & 0.727  & 0.445  \\ 
                                      & 1.0 & 0.4122 & 0.583 & 0.733  & 0.4505  \\ \hline 
 \multirow{3}{*}{RDR} & 0.5 & 0.3438 & 1.01 &  2.06   & 1.04   \\ 
                                        & 2.0 & 0.3491 & 1.01  &  2.06   & 1.04   \\ 
                                       & 1.0 & 0.3452 & 1  & 2  & 1   \\   
    \end{tabular}
\end{ruledtabular}
  \caption{Comparison between the critical point $\lambda_c$
  and  critical exponents $\beta$, $\nu_{\perp}$ and $\delta$ for non-Markovian (NM) SIS dynamics
  with different exponent $\alpha_I$, and the Markovian case (for which $\alpha_I=1$),
  on different underlying network topologies, 2D lattice and RDR network.
  The critical point $\lambda_c$ in NM SIS dynamics is evaluated by means of Eq.~\eqref{eq:lambda_approx_res}.
   }
  \label{tab:crit_exp}
\end{table}

Figure~\ref{fig:panel_lambda_exact} shows the prevalence $\rho^{st}$ at the steady state 
as a function of $\lambda_{eff}$ and $\lambda_{app}$,
for different values of $\alpha_I$ and different network substrates:
a two dimensional lattice with periodic boundary conditions, an Erd\H{o}s R\'enyi (ER) graph with $\av{k}=8$,
 a random degree regular (RDR) network with $\av{k}=8$, and a scale-free (SF) network with exponent $\gamma=2.5$.
One can see that different curves of the prevalence,
corresponding to different forms of the infection inter-event time distribution
collapse onto one another when plotted as a function of $\lambda_{eff}$.
This result is particularly noteworthy since two infection processes with the same average infection time $\langle t_I \rangle$
but different forms of $\psi_I(t)$ are known to  behave very differently~\cite{van_mieghem_non-markovian_2013},
showing huge differences in the prevalence $\rho^{st}$ for the same average infection time.
This is particularly true in the case of highly heterogeneous processes,
such as the one controlled by $\alpha_I = 0.25$,
with a very skewed form of the inter-event time distribution $\psi_I(t)$,
and by $\alpha_I = 10$, which corresponds to an almost-periodic process.

The curves plotted as functions of the approximate infection rate $\lambda_{app}$ are also almost indistinguishable  from the others,
showing that $\lambda_{app}$ is a very accurate approximation of the exact effective rate,
for every underlying network topology.
In the SI we also show that $\lambda_{app}$ is considerably different
 from the mean-field approximation proposed in~\cite{van_mieghem_non-markovian_2013},
 and far more accurate in describing extreme cases such as $\alpha_I = 0.25$ and $\alpha_I = 10$,
 see Supplementary Fig. \ref{fig:comparison_cator}.
 Interestingly, as we show in the SI, Eq.~\eqref{eq:lambda_approx_res}
is expected to converge to $\lambda_{eff}$ in the limit of low prevalence $\rho^{st} \ll 1$ and,
thus, close to the epidemic threshold, $\lambda_c$.
This implies that the exact critical point $\lambda_c$ of the non-Markovian SIS dynamics can be evaluated by
means of Eq.~\eqref{eq:lambda_approx_res}.
Using the same argument, we also conclude that the set of critical exponents of the non-Markovian dynamics
are the same as those of the Markovian one.

We check our hypothesis and evaluate the behavior of the non-Markovian SIS dynamics and its critical properties
by performing a finite size scaling (FSS) analysis.
We obtain the epidemic threshold $\lambda_{c}$, evaluated by means of Eq.~\eqref{eq:lambda_approx_res},
  and the set of critical exponents $\beta$, $\nu_{\perp}$ and $\delta$
for a non-Markovian SIS dynamics with $\alpha_I=0.5$ and  $\alpha_I=2$, on top of two-dimensional lattice and degree regular networks,
   by means of the lifespan method proposed in Ref.~\cite{PhysRevLett.111.068701},
see SI and Supplementary Fig. \ref{fig:FSS}.
 Table~\ref{tab:crit_exp} shows that the critical point and exponents
 for these cases 
  are in very good agreement with corresponding ones known in literature for Markovian SIS dynamics. 

In conclusion, we have demonstrated that non-Markovian SIS dynamics on arbitrary network topologies can be understood in terms of equivalent Markovian dynamics on the same substrates. This simplification of the temporal nature of discrete-state processes promises to find application in the wide variety of areas where non-Markovian aspects are recognized as increasingly influential.

\begin{acknowledgments}
We acknowledge support from the James S. McDonnell Foundation; the ICREA Academia prize, funded by the Generalitat de Catalunya; the MINECO projects no.~FIS2013-47282-C2-1-P and FIS2016-76830-C2-2-P; Generalitat de Catalunya grant no.~2014SGR608; and Science Foundation Ireland grant no.~11/PI/1026.
\end{acknowledgments}

\bibliographystyle{apsrev4-1}
\bibliography{Bibliography}

\clearpage

\begin{widetext}

\section{Supplementary Information}

\section{Derivation of the differential equation for the $n-$point correlation function}

The average of Eq (1) of the main text given the state of the system at time $t$, $\mathbf{n}(t) \equiv (n_1(t),n_2(t), \cdots, n_N(t))$, can be written as
\begin{equation}
\langle n_i(t + dt) | \mathbf{n}(t) \rangle = n_i(t) + dt A_i(t), \qquad A_i(t) = - n_i(t)\delta[t_i(t)]+(1-n_i(t)) \sum_j a_{ij} \lambda[\tau_{ji}(t)] n_j(t).
\label{eq:master_SI}
\end{equation}
Let us consider a set of $n$ nodes $\mathcal{I} \equiv \{i_1, i_2, \cdots i_n  \}$.
The correlation function between these $n$ nodes reads
\begin{equation}
\dot{\rho}_{i_1\cdots i_n}(t)  = \frac{1}{dt}\left \langle \left[  \prod_{i \in \mathcal{I}} \langle n_i(t + dt)| \mathbf{n}(t) \rangle -  \prod_{i \in \mathcal{I}} n_i(t) \right] \right \rangle,
\label{eq:npoint_SI}
\end{equation}
where the outer average is over the state of the system at time $t$. Notice also that the factorization in the first term of this equation is a direct consequence of the independence of the random variables $\xi_i$ and $\eta_i$ in Eq. (1) of the main text for different nodes. The first term in Eq~\eqref{eq:npoint_SI} can be written, by means of Eq~\eqref{eq:master_SI} as % \in Comb(k, \mathcal{I})}
\begin{equation}
 \prod_{i \in \mathcal{I}} \langle n_i(t + dt) | \mathbf{n}(t) \rangle = \sum_{k=0}^n \sum_{ \{ \mathcal{I}_k \} }
\left( dt \right)^{k}  \prod_{i \in \mathcal{I}_k}  n_i(t)
 \prod_{l \in \mathcal{I}_k}  A_l (t),
\end{equation}
where $\{ \mathcal{I}_k \}$ is the set of all subsets of  $\mathcal{I}$  containing $k$ nodes.
%a subset of  $\mathcal{I}$ and the sum runs over all the subsets of $\mathcal{I}$ containing $k$ nodes,
%$Comb(k, \mathcal{I})$.
Because of the term $dt^{k}$, however, the expansion to linear order in $dt$ is a reduced sum over $k < 2$, and thus
\begin{equation}
 \prod_{i \in \mathcal{I}} \langle n_i(t + dt) | \mathbf{n}(t) \rangle =
 \prod_{i \in \mathcal{I}}  n_i(t) +   dt \sum_{i \in \mathcal{I}} A_i(t) \prod_{k \in \mathcal{I}\setminus i } n_k(t).
 \end{equation}
Therefore the $n-$point correlation function reads
\begin{eqnarray}
\dot{\rho}_{i_1\cdots i_n}(t) & = &
\av{ \sum_{i \in \mathcal{I}} (1-n_i(t)) \sum_j a_{ij} \lambda[\tau_{ji}(t)]  n_j(t) \prod_{k \in \mathcal{I}\setminus i } n_k(t)
 - \sum_{i \in \mathcal{I}}  n_i(t) \delta(t_i(t))   \prod_{k \in \mathcal{I}\setminus i } n_k(t)  } \\
 & = &   \sum_{i \in \mathcal{I}} \left \langle \left[ (1-n_i(t)) \sum_j a_{ij} \lambda[\tau_{ji}(t)]  n_j(t)
  - n_i(t) \delta(t_i(t)) \right] \prod_{k \in \mathcal{I}\setminus i } n_k(t)  \right \rangle ,
 \end{eqnarray}
from which Eq (6) of the main text follows immediately.

\section{Time $\tau_{ij}$ of active link $i-j$ does not depend on the states of other nodes different from $i$ and $j$}

A critical step in our approach is to prove that
\begin{equation}
\text{Prob}(\tau_{ji};t |  n_i = 0, n_j = 1, \{n_k=1, k \in \mathcal{I}_i\}) = \text{Prob}(\tau_{ji} ;t | n_i = 0,  n_j = 1).
\label{tauji}
\end{equation}
The probability in the left hand side of this equation can be written as
\begin{equation}
\text{Prob}(\tau_{ji};t |  n_i = 0, n_j = 1, \{n_k=1, k \in \mathcal{I}_i\})=\int \cdots \int \phi(\tau^R_i,\tau^I_j, \{\tau^I_k\};t) \phi(\tau_{ji}|\tau^R_i,\tau^I_j, \{\tau^I_k\})  d \tau^R_i d\tau^I_j \prod_{k \in \mathcal{I}_i} d \tau^I_k,
\end{equation}
where $\phi(\tau^R_i,\tau^I_j, \{\tau^I_k\};t)$ is the joint probability density, at time $t$, that given that node $i$ is susceptible and nodes $j$ and $ \{k \in \mathcal{I}_i\}$ are infected, the time elapsed since $i$ recovered is $\tau^R_i$ and the times elapsed since $j$ and $ \{ k \in \mathcal{I}_i\}$ became infected are $\tau^I_j$ and $\{\tau^I_k\}$, respectively. By  Bayes' rule, $\phi(\tau_{ji}|\tau^R_i,\tau^I_j, \{\tau^I_k\})$ is the probability density of the time $\tau_{ji}$ conditioned on the times $\tau^R_i,\tau^I_j, \{\tau^I_k\}$. However, it is easy to see that since infection events take place in active links independently, once $\tau^R_i$ and $\tau^I_j$ are fixed, $\tau_{ji}$ is totally independent of the elapsed times since nodes other than $j$ became infected. Therefore,
\begin{equation}
\phi(\tau_{ji}|\tau^R_i,\tau^I_j, \{\tau^I_k\})=\phi(\tau_{ji}|\tau^R_i,\tau^I_j),
\label{phi_tauji}
\end{equation}
which directly gives
%from where it follows immediately 
the result in Eq.~\eqref{tauji}.

\section{General formalism for $\lambda_{eff}$}

%\Note{This Section (until Eq.~\eqref{eq:expon_ansatz}) is devoted only to find $\lambda_{mf}$ (by Cator et al) as a special case.
%Is it necessary?}
Using the result in Eq.~\eqref{phi_tauji}, at the steady state the probability density $ \phi (\tau_{ji})$ of the time elapsed since the infection process of node $j$ to node $i$ started,
given that node $i$ is susceptible and node $j$ is infected,
%the time elapsed since the infection process of node $j$ to node $i$ started is equal to $\tau_{ji}$ 
can be written in general as
\begin{equation}
 \phi (\tau_{ji}) = \int \int \phi(\tau_{ji} | \tau^I_j, \tau^R_i) \phi(\tau^I_j, \tau^R_i) d\tau^I_j d\tau^R_i,
 \label{eq:tau_ji_exact}
\end{equation}
where $\phi(\tau^I_j, \tau^R_i)$ is the joint probability that the time elapsed since $j$ became infected is equal to $\tau^I_j$ and the time elapsed since $i$ recovered is equal to $\tau^R_i$.
%JG One can
%If one neglects the correlations between $\tau^I_j$ and $\tau^R_i$,
%We measure the correlation between $\tau^I_j$ and $\tau^R_i$ \JG{in numerical simulations},  and find that the processes are weakly correlated \JG{{\bf[JG: Maybe add a plot for evidence of this?]}}.
If we assume that the two process are uncorrelated,
$\phi(\tau^I_j, \tau^R_i)$ can be factorized into $\phi(\tau^I_j, \tau^R_i) = \phi_I(\tau^I_j) \phi_R(\tau^R_i)$,
and Eq. \eqref{eq:tau_ji_exact} reduces to
\begin{equation}
\phi(\tau_{ji}) = \int_0^\infty d\tau^I_j \phi_I(\tau^I_j) \int_0^\infty d\tau^R_i \phi_R(\tau^R_i) \Big\{
\Theta(\tau^I_j - \tau^R_i) \phi(\tau_{ji} | \tau^R_i \leq \tau^I_j) +  \Theta( \tau^R_i - \tau^I_j) \phi(\tau_{ji} | \tau^R_i > \tau^I_j) \Big\} ,
\label{eq:phi_tau_exact_cond }
\end{equation}
where $\Theta(t)$ is the Heaviside step function,
$\phi_I(\tau^I_j)$ is the probability that the time elapsed since $j$ became infected is equal to $\tau^I_j$ and
 $\phi_R(\tau^R_i)$ is the probability that the time elapsed since $i$ recovered is equal to $\tau^R_i$.
The conditional probability $\phi(\tau_{ji} | \tau^R_i > \tau^I_j) $ is simply $\phi(\tau_{ji} | \tau^R_i > \tau^I_j) = \delta(\tau_{ji} - \tau^I_j)$, and
\begin{equation}
\phi(\tau_{ji} | \tau^R_i \leq \tau^I_j) = \int_0^\infty \Theta(\tau^I_j-\tau^R_i-\tau) \delta(\tau_{ji} - (\tau^I_j-\tau)) \Psi_I(\tau^I_j-\tau^R_i-\tau)
\sum_{n=0}^\infty P_n(\tau) d\tau,
\end{equation}
where $n$ is the number of infection attempts of node $j$ to node $i$,
$P_n(\tau) $ is the probability that the time elapsed since node $j$ became infected and the moment of his $n$-th fire is equal to $\tau$, 
and  $\Psi_I(\tau^I_j - \tau^R_i - \tau)$ is the probability that the time elapsed between the $n$-th fire and the $n+1$-th fire
is greater than $\tau_{ji}- \tau^R_i$.
As computed in Eq. (4) of the main text, the probability that the time elapsed since $j$ became infected is equal to $\tau^I_j$ is simply
\begin{equation}
\phi_I(\tau^I_j) = \tilde{\delta} \Psi_R(\tau^I_j).
\label{eq:phi_tau_0}
\end{equation}
The survival probability $\Psi_R(\tau^I_j)$ of recovery events can be written as
\begin{equation}
\label{eq:omega}
 \Psi_R(\tau^I_j) = \int_0^\infty \omega(u) e^{-u\tau^I_j} du = \widehat{\omega}(\tau^I_j),
 \end{equation}
where  $\omega(u)$ is the inverse Laplace transform of $\Psi_R(\tau^I_j)$. % $\omega(u) \equiv \mathcal{L}^{-1} \{ \Psi_R(t) \}$.
 In Laplace space, the probability distribution $P_n(\tau)$ has a convenient form, $\widehat{P}_n(u) = \Big[ \widehat{\psi}_I(u) \Big]^n$,
 where $\widehat{ \psi}_I(u)$ is the Laplace transform of $\psi_I(t)$.
By inserting Eqs. \eqref{eq:phi_tau_0} and \eqref{eq:omega} into Eq. \eqref{eq:phi_tau_exact_cond }
%JG one can
we obtain
\begin{equation}
\phi(\tau_{ji}) = \tilde{\delta} \int_0^\infty d\tau^R_i \phi_R(\tau^R_i) \int_0^\infty du\, e^{-u \tau_{ji}} \omega(u)
\left\{ \theta (\tau^R_i - \tau_{ji}) + \theta (\tau_{ji} - \tau^R_i ) \Psi_I(\tau_{ji} - \tau^R_i)  \frac{1}{1- \widehat{\psi}_I (u)} \right\} .
\label{eq:tau_ij_exact}
\end{equation}
 By inserting the form of $\phi(\tau_{ji})$ into Eq. (10) of the main text,
  we obtain an expression for the infection rate $\lambda_{eff}$
\begin{equation}
\label{eq:lambda_exact}
\lambda_{eff} = \int_0^\infty du \, \omega(u) \left \{  \widehat{ \left[ \lambda_I \Phi_R  \right]}(u) +
   \reallywidehat{  \left[  \lambda_I \left[\phi_R * \Psi_I \right] \right] } (u) \frac{1}{1-  \widehat{\psi}_I(u)} \right\},
\end{equation}
where $\lambda_I$ is the infection hazard rate,
$\Phi_R(\tau^R_i)$ and  $\Psi_I(t)$ are the survival probabilities of $\phi_R(\tau^R_i)$ and  $\psi_I(t)$, respectively and
$\phi_R * \Psi_I $ is the convolution between $\phi_R(\tau^R_i)$ and $\Psi_I(t)$.
At this point, some ansatz regarding the form of $\phi_R(\tau^R_i)$, the probability that the time elapsed since $i$ recovered is equal to
$\tau^R_i$, is needed to continue.
We note that if one does not consider the state of node $i$ in the probability $\phi_R(\tau_{ji})$,
which corresponds to inserting $\phi_R(\tau^R_i) = \delta(\tau^R_i)$ into Eq. \eqref{eq:tau_ij_exact},
one obtains
\begin{equation}
\lambda_{mf} = \tilde{\delta} \int_0^{\infty} \omega(u) \frac{\widehat{ \psi}_I(u) }{1- \widehat{ \psi}_I(u)}du.
\label{eq:lambda_cator}
\end{equation}
This effective infection rate $\lambda_{mf}$, already found in Cator et al.~\cite{PVM_nonMarkovianSIS_NIMFA_2013} by using a mean field approximation,
is now obtained within a more general formalism. %, which highlights its validity and
A different possibility is to consider $\phi_R(\tau^R_i) $ equal to an exponential distribution,
\begin{equation}
\phi_R(\tau^R_i) = w e^{-w\tau^R_i},
\label{eq:expon_ansatz}
\end{equation}
with rate $w$.
The rate $w$ can be written as a simple function of the prevalence $\rho$, $w= \delta \rho / (1-\rho)$.
However, even if it were not possible to find a closed analytic form for the effective infection rate,
one can resort to numerical simulation in order to compute $\lambda_{eff}$.
One can see that the exponential ansatz for the form of $\phi_R(\tau^R_i)$ is correct for large values of the prevalence $\rho$,
but it fails for low prevalence, thus close to the epidemic threshold.

\section{Approximations to $\lambda_{app}$}

To find an infection rate which is accurate and analytically treatable close to the epidemic threshold, we follow a different approach. As stated in the main text, we consider here the probability density
$\psi (\tau_{ji}) \equiv \lim_{t\rightarrow \infty}   p(\tau_{ji}, n_i=0; t | n_j=1)$,
which is the join probability that, given that node $j$ is infected at the observation time,
node $i$ is susceptible and the time elapsed since the last infection attempt from $j$ to $i$ is equal to $\tau_{ji}$.
Our approximation consists of estimating the probability that node $i$ is susceptible at the observation time $t$,
which depends on the time instant at which node $i$ became infected, this time instant being unknown in principle.
\begin{figure}[tbp]
  \begin{center}
    \includegraphics[width=\linewidth,angle=0]{\FigPath/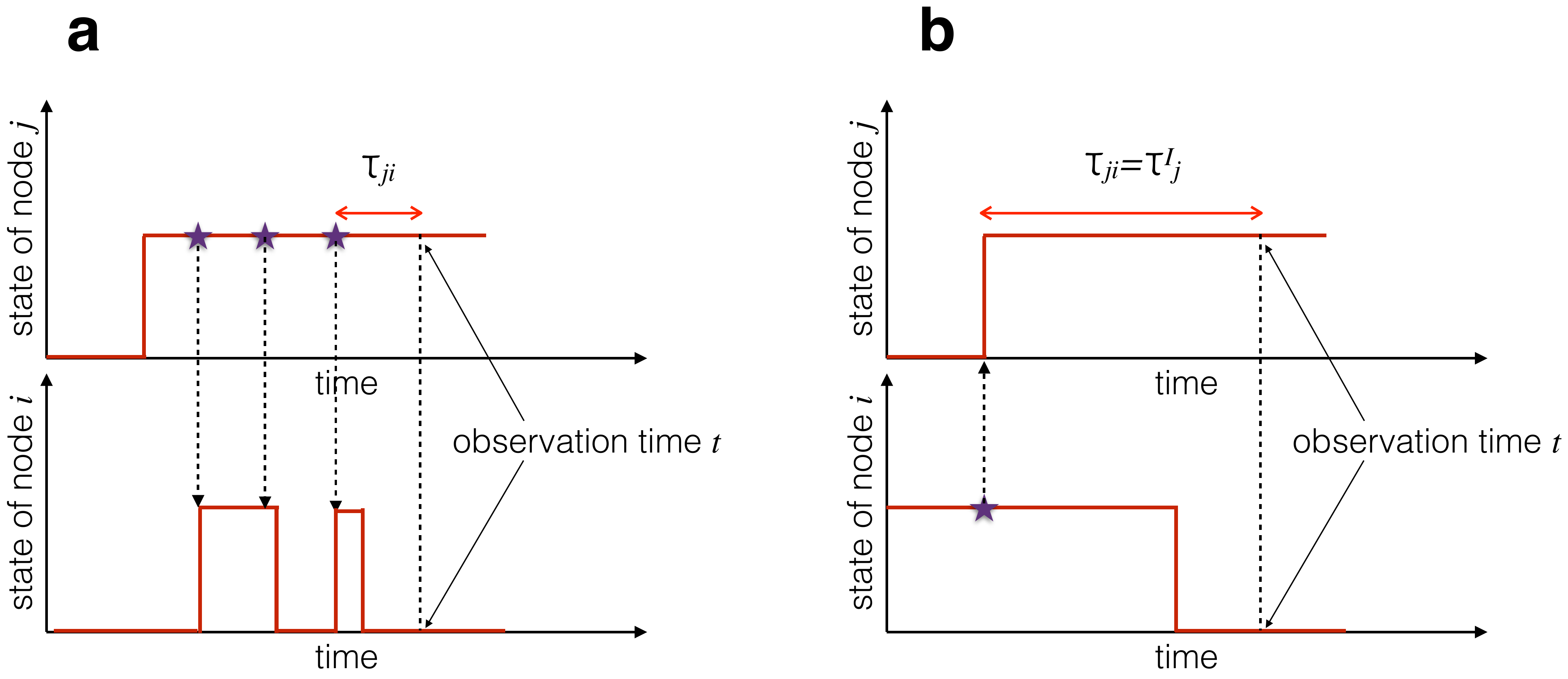}
  \end{center}
  \caption{Sketch of two possible ways to obtain the time $\tau_{ji}$ and, simultaneously, node $i$ infected at the observation time. In {\bf a}, node $j$ has attempted to infect $i$ at least once at time $t-\tau_{ji}$. This implies that node $i$ must necessarily be infected at that moment and, thus, it has to recover before the observation time. In {\bf b}, node $j$ has not attempted to infect $i$ since it became infected by node $i$. In this case, we know that $i$ was infected when $j$ became infected and, again, it has to recover before the observation time.}
  \label{fig:sketch3}
\end{figure}

We first %restrict to the case of Markovian recovery, so to take advantage of its memoryless property.
%Then we
 consider the case of low prevalence, $\rho^{st} \ll 1$.
Then, let us consider separately the cases in which node $j$ attempts at least once to infect node $i$, $n>0$,
and the case of no attempts, $n=0$.
In the first case, at the time of the last infection event from $j$ to $i$, $t-\tau_{ji}$, node $i$ either was already infected (in which case the infection attempt is ineffective) or it became infected by this event (see Fig.~\ref{fig:sketch3}a). In both cases, we are certain that node $i$ is in an infected state at time $t-\tau_{ji}$ and, thus, the probability that node $i$ recovers before the observation time $t$ is $1-\Psi_R(\tau_{ji})$. If the prevalence is low, the probability that node $i$ is subsequently infected by one of its neighbor (other than $j$) and then recovers before the observation time is also very low, and we assume it to be zero. With these assumption, %of at least one infection attempt from $j$ to $i$,
the probability that node $i$ is susceptible at time $t$ is simply $1 -  \Psi_R(\tau_{ji})$.
If node $j$ does not attempt to infect node $i$, we cannot know for certain the state of node $i$.
%In the limit of low prevalence, one might assume that node $i$ is susceptible at time $t-\tau^I_j$,
%and he remains in this state till time $t$,  so that $\text{Prob}(n_i=0,t | n=0) = 1$.
However, given that node $j$ became infected at time $t-\tau^I_j$, one of his neighbors must have infected him.
Let us consider that node $j$ has degree $k$. If the prevalence is low, it is very unlikely to find more than one neighbor of node $j$ infected simultaneously and we assume that only one of his neighbors was infected and infected him.
With probability $1/k$, such infected node is node $i$ (See Fig.~\ref{fig:sketch3}b), so that $i$ is infected at time $t-\tau^I_j$,
and the probability that node $i$ is then susceptible at the observation time $t$ is $1 -  \Psi_R(\tau^I_j)$.
 With probability $1-1/k$, the infected node is a neighbor other than $i$ and, thus, we assume that node $i$ was susceptible at time $t-\tau^I_j$ and it will remain in this state until time $t$ with probability equal to one.
Summing up, if node $j$ attempts at least once to infect node $i$, then the probability that it is susceptible at the observation time is $ 1 -  \Psi_R(\tau_{ji})$. Instead, if node $j$ does not attempt to infect node $i$,  this probability reads $ (1 -  \Psi_R(\tau^I_j))/k + (k-1)/k = (k - \Psi_R(\tau^I_j))/k$. In the limit of low prevalence, we expect this approximation to be exact. In the following, we also approximate the value of $k$ by the average degree, $\av{k}$.

The probability density $ \psi (\tau_{ji}) $ can be written as
\begin{equation}
\psi (\tau_{ji})   = \int_0^\infty \psi(\tau_{ji} | \tau^I_j) \phi_I(\tau^I_j) d\tau^I_j,
\label{eq:psi_tau_ij_corrected}
\end{equation}
where again $\phi_I(\tau^I_j)$ is the probability that the time elapsed since $j$ became infected is equal to $\tau^I_j$.
The conditional probability $ \psi(\tau_{ji} | \tau^I_j)$ is
\begin{equation}
 \psi(\tau_{ji} | \tau^I_j) =   \delta(\tau_{ji} - \tau^I_j ) \Psi_I(\tau^I_j )  \left[  \frac{\av{k} -\Psi_R(\tau^I_j )}{\av{k}} \right]  +
  \int_0^{\tau^I_j} \delta(\tau_{ji} - (\tau^I_j - \tau) )  \left[  1-\Psi_R(\tau_{ji} ) \right]
   \Psi_I(\tau^I_j - \tau)  \sum_{n=1}^\infty P_n(\tau)  d\tau
\end{equation}
where the first term accounts for the case in which there are no infection attempts from $j$ to $i$, $n=0$,
 while the second term accounts for the case $n > 0$.
%where we separated the case of no infection attempt from the case $n > 0$ and
%we recall that $\tau^I_j$ is the time elapsed at time $t$ since $j$ became infected and
%$\tau$ is the time elapsed since node $j$ became infected and his last infection attempt to node $i$.
%The first term accounts for the case $n=0$, where node $i$ is susceptible with probability equal to one.
By inserting Eq. \eqref{eq:phi_tau_0} into Eq. \eqref{eq:psi_tau_ij_corrected} and integrating over  $\tau^I_j$,
the probability $ \psi(\tau_{ji})$ reads
\begin{equation}
 \psi(\tau_{ji}) = \tilde{\delta} \Psi_I(\tau_{ji}) \left\{  \Psi_R(\tau_{ji})     \left[  \frac{\av{k}-\Psi_R(\tau_{ji} )}{\av{k}} \right]
        + \left[ 1- \Psi_R(\tau_{ji}) \right]   \int_0^{\infty} \Psi_R(\tau_{ji} + \tau)  \sum_{n=1}^\infty P_n(\tau) d\tau \right \} .
\end{equation}
If we restrict to the case of Markovian recovery, we can use its memoryless property, $\Psi_R(\tau_{ji} + \tau) = \Psi_R(\tau_{ji}) \Psi_R(\tau)$.
By using the convenient Laplacian form of the probability distribution $P_n(\tau)$, one can obtain
\begin{equation}
 \psi(\tau_{ji}) = \tilde{\delta} \frac{ \Psi_I(\tau_{ji}) \Psi_R(\tau_{ji}) } {1 - \widehat{\psi}_I ( \tilde{\delta}) }
 \left \{ 1 - \Psi_R(\tau_{ji}) \left[ \av{k}^{-1} \left( 1- \widehat{\psi}_I ( \tilde{\delta}) \right) +   \widehat{\psi}_I ( \tilde{\delta}) \right] \right \}  .
 \label{eq:tau_ji_final}
\end{equation}
The normalization of $ \psi(\tau_{ji})$ reads
\begin{equation}
 \label{eq:norm_tau_ji}
 \mathcal{N} = \int_0^\infty \psi(\tau_{ji}) d\tau_{ji} =  \frac{ \tilde{ \delta} } {1 - \widehat{\psi}_I ( \tilde{\delta}) }
 \left \{   \widehat{\Psi}_I ( \tilde{\delta}) -  \widehat{\Psi}_I (2 \tilde{\delta})
 \left[ \av{k}^{-1} \left( 1- \widehat{\psi}_I ( \tilde{\delta}) \right) +   \widehat{\psi}_I ( \tilde{\delta}) \right] \right \}  ,
 \end{equation}
 therefore, by inserting  Eq.~\eqref{eq:tau_ji_final} and Eq.~\eqref{eq:norm_tau_ji} into  Eq. (11) of the main text,
one finally obtains the approximate infection rate $\lambda_{app}$ presented in Eq. (12) of the main text.

In Fig.~\ref{fig:comparison_cator}, we show the steady-state prevalence  $\rho^{st}$  as a function of the approximate effective infection rate $\lambda_{app}$, given by Eq. (12) of the main text, and the mean field effective rate $\lambda_{mf}$ given by Eq.~\eqref{eq:lambda_cator} for two extreme values of the exponent $\alpha_I$ controlling the interevent time infection distribution, $\alpha_I=0.25$ and $\alpha_I=10$. One can see that the curves for $\lambda_{mf}$ do not collapse onto one another, especially for the lattice and SF network substrate.

\begin{figure}[tbp]
  \begin{center}
    \includegraphics[width=12cm,angle=0]{\FigPath/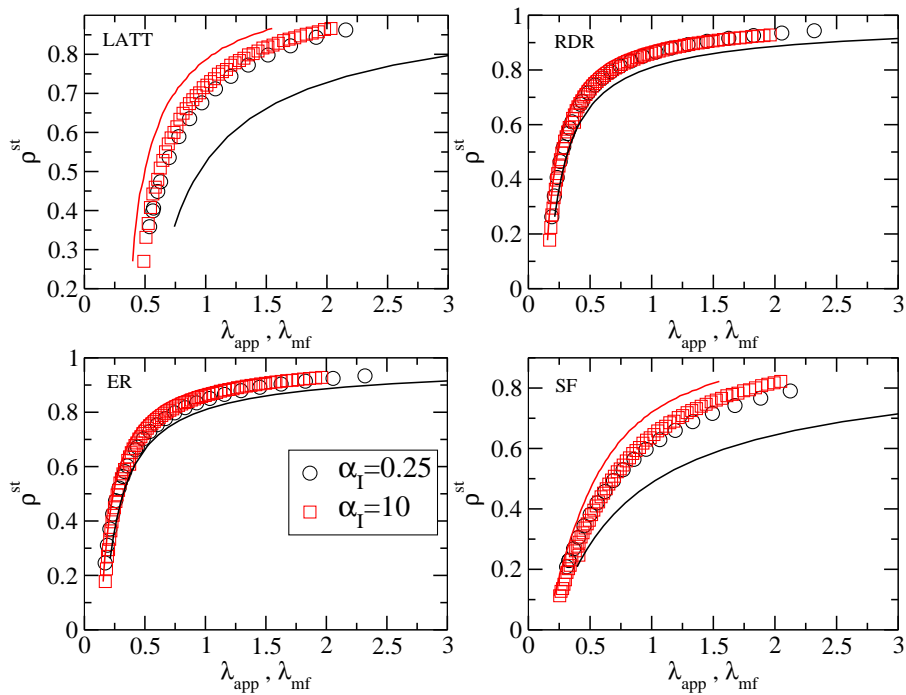}
  \end{center}
  \caption{
  Prevalence $\rho$ as a function of the approximate effective infection rate $\lambda_{app}$ (points)
  and the mean field effective rate $\lambda_{mf}$ (continuous line),
   	for different values of the exponent $\alpha_I$ and different network substrate.}
  \label{fig:comparison_cator}
\end{figure}

\section{Numerical simulations of the non-Markovian SIS dynamics}

To check the validity of the effective infection rates $\lambda_{eff}$ and $\lambda_{app}$,
we run extensive numerical simulations of the non-Markovian SIS dynamics.
For each value of the average infection time $\av{t_I}$ and fixed average recovery time $\av{t_R}= \tilde{\delta}^{-1} =1$,
we simulate the non-Markovian SIS dynamics by implementing an algorithm based on a queue of infection and recovery events.

At time $t=0$, all nodes are in a susceptible state, and a set of $fN$ randomly chosen  nodes, with $f=0.5$,
change their state to the infected one.
In the algorithm, whenever a node $i$ changes his state from susceptible to infected at time $t$,
he first randomly extracts his recovery time $t_R$ from the distribution $\psi_R(t)$,
and pushes his recovery event at time $t+t_R$ to the queue.
He also starts $k$ independent infection processes to his $k$ neighbors.
In each infection process to a neighbor $j$,
an infection event from node $i$ to node $j$ is scheduled at time $t+t_I^1$,
where  $t_I^1$ is randomly extracted from the distribution $\psi_I(t)$, only if $t_I^1 < t_R$,
that is if node $i$ is still infected at time $t+t_I^1$.
A second infection event from node $i$ to node $j$  is scheduled at time $t+t_I^1+ t_I^2$,
where $t_I^2$ is randomly extracted from the distribution $\psi_I(t)$, only if $t_I^1 + t_I^2 < t_R$,
and so on until $n$ (with possibly $n=0$) infection events are generated and pushed to the queue.

The queue is pulled by following the time order of the events.
If the pulled event is the recovery of node $i$, $i$ changes his state from infected to susceptible.
If the pulled event is an infection event from node $i$ to node $j$, and $j$ is already in a infected state, nothing happens,
otherwise node $j$ changes his state from susceptible to infected and schedules his recovery and infection events, pushing them to the queue.
The queue is pulled until either no more events are left (and so all nodes are susceptible)
or the time reaches a time $T_{max}$, set conveniently.
In order to measure the prevalence in the steady state $\rho^{st}$ and the effective infection rate $\lambda_{eff}$,
we sample $N_{s} = 10^4$ time instants uniformly chosen in $[T_{min}, T_{max}]$,
with $T_{min}$ chosen such that the stationary state is reached long before it.
For each time instant, we measure the prevalence and the values of $\tau_{ij}$ for each active link between nodes $i$ and $j$,
so as to calculate $\lambda_{eff}$ by means of Eq. (10) of the main text.

We have double checked our event queue algorithm by simulating the non-Markovian SIS dynamics
with a non-Markovian Gillespie algorithm \cite{boguna_simulating_2013},
which is much  slower, and we obtained identical results for the prevalence and the effective infection rate.

\section{Epidemic threshold and critical exponents}

We run extensive numerical simulations of the non-Markovian SIS dynamics
in order to evaluate its behavior close to the epidemic threshold.
We consider $\alpha_I = 0.5$ and $\alpha_I = 2$, and two different network substrates,
2D lattice and RDR network. We address the critical properties by means of the lifespan method \cite{PhysRevLett.111.068701},
in which the infection starts with a single infected node.
In the lifespan method, each realization is characterized by its lifetime, $T$, and its coverage, $C$,
  defined as the number of distinct nodes that have become infected at least once.
  We let each realization run until either the coverage $C$ reaches a certain threshold $C^*$
  (and we consider it endemic), or the realization dies out, and we measure its lifetime $T$.
We set $C^* = \Theta N$, with $\Theta = 0.9$.
We then measure the probability of having an endemic realization $P$, the average lifetime $\av{T}$ and average square lifetime $\av{T^2}$
over a number of runs $N_{run}$, as a function of the average infection time $\av{t_I}$
(corresponding to an effective rate $\lambda_{app}$, hereafter $\lambda$ for brevity) close to the epidemic threshold,
for different sizes $N$. We set $N_{run} =10^5$ for lattice, $N_{run} =10^6$ for RDR networks.
For each value of $N$,  $\lambda$, $\alpha_I$ and network substrate we fit the curves of
$\av{T}$ and $\av{T^2}$ to find the peaks $\av{T}_p$ and $\av{T^2}_p$ and their corresponding values of $\lambda_p^1$
and $\lambda_p^2$. We set $\lambda_p$ as the average of $\lambda_p^1$ and $\lambda_p^2$,
provided that $\lambda_p$ falls within the $\lambda_p^1$ and $\lambda_p^2$ standard errors.
The corresponding endemic probability at the peak $P_p$ is interpolated from the data.

The set of equations we used to evaluate the critical point $\lambda_c $ and critical exponents $\beta, \delta, \nu_{\perp}$ are
\begin{eqnarray}
P(\lambda_c,N) & \sim & N^{-\beta/\nu_{\perp}} \label{eq:P(l_c,N)}\\
P_p(N) &\sim &N^{-\beta/\nu_{\perp}} \label{eq:P(N)} \\
|\lambda_c - \lambda_p(N)| &  \sim & N^{1/\nu_{\perp}}. \label{eq:l_c(N)} \\
\av{T^n}_p(N) &\sim & N^{\gamma_n/\nu_{\perp}} \label{eq:T_N)} \\
\end{eqnarray}
We first evaluate the critical threshold $\lambda_c$ by plotting the endemic probability $P(\lambda,N)$ as a function of $N$,
for several values of $\lambda$ close to $\lambda_c$, see the first row of Fig. \ref{fig:FSS}.
%For $\lambda = \lambda_c$, $P(\lambda_c,N)$ is expected to be a power-law with exponent equal to $\beta/\nu_{\perp}$,
Through Eq~\eqref{eq:P(l_c,N)}, we estimate the value of $\lambda_c$ to be the one which produces
the best fit of  $P(\lambda_c,N)$ as a power-law.
    We then plot the endemic probability at the peak $P_p$ as a function of the size $N$, see the second row of Fig. \ref{fig:FSS},
 %  $P_p(N)$ is expected to be a power-law with exponent equal to $\beta/\nu_{\perp}$.
   and the difference $|\lambda_c - \lambda_p|$ as a function of $N$, see the third row of Fig. \ref{fig:FSS}.
   %   $|\lambda_c - \lambda_p|(N)$ is expected to be a power-law with exponent equal to $1/\nu_{\perp}$,
       We estimate $\nu_{\perp}$ by means of Eq~\eqref{eq:l_c(N)}.
        By means of Equations~\eqref{eq:P(l_c,N)} and~\eqref{eq:P(N)} we estimate $\beta/\nu_{\perp}$ equal to
 the average of the fits of $P(\lambda_c,N)$ and $P_p(N)$,
   provided that $\beta/\nu_{\perp}$ falls within the standard errors of the two fits,
   and so we calculate $\beta$, knowing the value of $\nu_{\perp}$.
       Finally, we plot the height of the peaks $\av{T}_p$ and $\av{T^2}_p$ as a function of $N$, see the fourth row of Fig. \ref{fig:FSS}.
       We estimate $\gamma_2$ and $\gamma_1$ for lattices and $\gamma_2$ for RDR networks
        by means of Eq~\eqref{eq:T_N)}, knowing the value of $\nu_{\perp}$.
        For lattices, we calculate $\delta$ by means of the equivalence $\gamma_n \sim n - \delta$ while
         for RDR networks we first check that $\av{T}_p$ diverges logarithmically as a function of $N$
         and then we calculate $\delta$ by means of the equivalence $\gamma_n  = n - \delta$.
        The results are reported in Table 1 of the main text.

\begin{figure}[tbp]
  \begin{center}
    \includegraphics[width=12cm,angle=0]{\FigPath/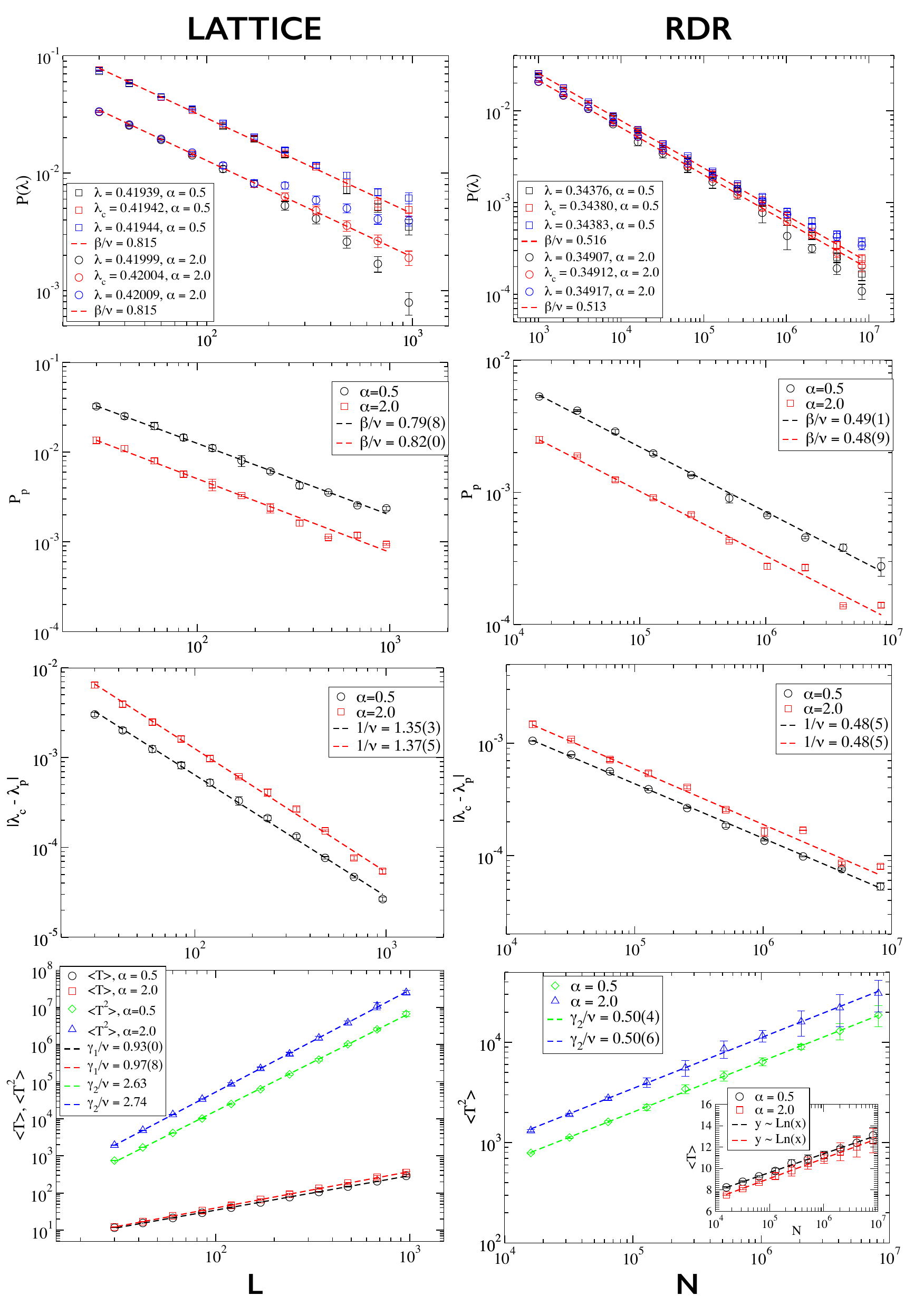}
  \end{center}
  \caption{
  Finite size scaling of a non-Markovian SIS dynamics with $\alpha_I=0.5$ and  $\alpha_I=2$,
  on 2D lattice (on the left) and RDR network (on the right).
  Symbols represent the results of numerical simulations, dashed lines represent power-law
  (or logarithmic, in the case of $\av{T}$ for RDR) fits.
  In this Figure, we refer to $\lambda_{app}$ as to $\lambda$ for brevity. Notice that, to compare with the values found in the literature, in the case of the lattice we use the side of the lattice $L$ instead of the number of nodes $N=L^2$.
  Plots show, from first to last row:
  (1) Probability that an outbreak is endemic for different values of $\lambda$, $P(\lambda)$, as a function of the size $N$.
  %For $\lambda = \lambda_c$, $P(\lambda_c,N)$ is expected to be a power-law with exponent equal to $\beta/\nu_{\perp}$,
   % providing us with a way to evaluate $\lambda_c$.
   (2) Probability that an outbreak is endemic, $P(\lambda_p)$, as a function of the size $N$,
   for $\lambda_p$ corresponding to the peak of $\av{T^2}$.
   (3) Difference $|\lambda_c - \lambda_p|$ as a function of $N$, for $\lambda_p$ corresponding to the peak of $\av{T^2}$.
      (4) Peak of $\av{T}$ and $\av{T^2}$ as a function of $N$.}
  \label{fig:FSS}
\end{figure}

\end{widetext}

\end{document}